\documentclass[12pt]{iopart}

\pdfoutput=1

\usepackage{iopams}  
\usepackage{graphicx}
\usepackage{setstack}
\usepackage{color}
\usepackage{cite}

\usepackage{braket}

\newcommand{\eqref}[1]{(\ref{#1})}
\usepackage{bm}
\usepackage{wasysym} 

\DeclareMathAlphabet      {\mathitbf}{OML}{cmm}{b}{it}

\usepackage{array}

\usepackage{stackrel}
\newcolumntype{P}[1]{>{\centering\arraybackslash}p{#1}}
\newcommand{\leftrarrows}{\mathrel{\raise.75ex\hbox{\oalign{%
  $\scriptstyle\leftarrow$\cr
  \vrule width0pt height.5ex$\hfil\scriptstyle\relbar$\cr}}}}
\newcommand{\lrightarrows}{\mathrel{\raise.75ex\hbox{\oalign{%
  $\scriptstyle\relbar$\hfil\cr
  $\scriptstyle\vrule width0pt height.5ex\smash\rightarrow$\cr}}}}
\newcommand{\Rrelbar}{\mathrel{\raise.75ex\hbox{\oalign{%
  $\scriptstyle\relbar$\cr
  \vrule width0pt height.5ex$\scriptstyle\relbar$}}}}
\newcommand{\longleftrightarrows}{\leftrarrows\joinrel\Rrelbar\joinrel\lrightarrows}

\begin{document}

\title[KPZ universality class for the critical dynamics of reaction-diffusion fronts]{Kardar-Parisi-Zhang universality class for the critical dynamics of reaction-diffusion fronts}

\author{B. G. Barreales$^{1,2}$, J. J. Mel\'endez$^{1,2}$, R. Cuerno$^3$, J. J. Ruiz-Lorenzo$^{1,2}$}

\address{$^1$ Departamento de  F\'{\i}sica, Universidad de Extremadura, 06006 Badajoz, Spain}
\address{$^2$ Instituto de Computaci\'on Cient\'{\i}fica Avanzada de Extremadura (ICCAEx), Universidad de Extremadura, 06006 Badajoz, Spain}
\address{$^3$ Departamento de Matem\'aticas and Grupo Interdisciplinar de Sistemas Complejos (GISC), Universidad Carlos III de Madrid, 28911 Legan\'es, Spain}

\date{\today}

\begin{abstract}
We have studied front dynamics for the discrete $A+A \leftrightarrow A$ reaction-diffusion system, which in the continuum is described by the (stochastic) Fisher-Kolmogorov-Petrovsky-Piscunov equation. We have revisited this discrete model in two space dimensions by means of extensive numerical simulations and an improved analysis of the time evolution of the interface separating the stable and unstable phases. In particular, we have measured the full set of critical exponents which characterize the spatio-temporal fluctuations of such front for different lattice sizes, focusing mainly in the front width and correlation length. These exponents are in very good agreement with those computed in [E.\ Moro, Phys.\ Rev.\ Lett.\ {\bf 87}, 238303 (2001)] and correspond to those of the Kardar-Parisi-Zhang (KPZ) universality class for one-dimensional interfaces. Furthermore, we have studied the one-point statistics and the covariance of rescaled front fluctuations, which had remained thus far unexplored in the literature and allows for a further stringent test of KPZ universality.
\end{abstract}

\maketitle 

\section{Introduction}
\label{sec:intro}

Reaction-diffusion systems stand out as a context for collective behavior and nonlinear properties within statistical mechanics \cite{cross2009}, with experimental instances ranging from chemical reactions \cite{epstein1998}, to developmental biology \cite{tsimring2014}, to epidemic processes in complex networks \cite{pastor-satorras2015}, or reactive turbulence \cite{boffetta2017}.
Historically, a major role in the understanding of this class of systems has been played by the Fisher-Kolmogorov-Petrovsky-Piscunov (FKPP) equation \cite{fisher1937,kolmogorov1937} for a non-negative scalar field $\rho(\bm{r},t)$, e.g.\ a population density, which couples a diffusion process to logistic growth as
\begin{equation}
    \partial_t \rho=D\nabla^2 \rho + \rho - \rho^2 ,
    \label{eq:fisher}
\end{equation}
where $D>0$ is a parameter. Here, $\bm{r} = (r_{\bot},\bm{r}_{\parallel}) \in \mathbb{R}^{d}$, with $r_{\bot} \in \mathbb{R}$ and $\bm{r}_{\parallel} \in \mathbb{R}^{d_{\parallel}}$ so that $d=d_{\parallel}+1$. As is well known \cite{murray2002}, for initially segregated conditions, e.g.\ $\rho(\bm{r},t=0) = \rho_{\rm eq}$ for $r_{\bot}\leq 0$ and $\rho(\bm{r},t=0) = 0$ for $r_{\bot}> 0$, Eq.\ \eqref{eq:fisher} describes the propagation of a front $h(\bm{r_{\parallel}},t)$ along the $r_{\bot}$ coordinate, separating a stable phase ($\rho=1$) which invades a marginally unstable one $(\rho=0)$ \cite{saarloos2003}. Again, examples abound, from DNA reaction networks \cite{zadorin2015} to colloidal systems \cite{tanaka2016,zadorin2017}, to the epidemic spread of diseases \cite{sun2016}, or the dynamics of invasions among human populations \cite{hui2017}.

The description of front propagation provided by the FKPP equation is accurate at a mean-field (MF) level, but it is natural to refine it by exploring the effect of stochastic fluctuations, e.g.\ in the population density. Actually, fluctuations in reaction-diffusion systems not only just quantitatively modify system properties, as e.g.\ the average front velocity \cite{sagues2007,hallatschek2011}; they can even lead to novel phenomena by themselves, like the emergence of system configurations which are not available within a MF approximation \cite{sagues2007,tauber2014}, or they can dominate the large-scale behavior of the system \cite{tauber2014,nesic2014}. A natural approach to account for fluctuations is to resort to more microscopic, discrete models \cite{odor2004} whose macroscopic evolution is consistent with the front dynamics dictated by Eq.\ \eqref{eq:fisher} \cite{odor2004,panja2004}, but which allow for explicit assessment of the dynamical role of external or internal noise. Thus for instance, the $A+A \leftrightarrow A$ reaction-diffusion model \cite{ben-avraham1990} has been shown \cite{pechenik1999,doering2003} to implement a stochastic generalization of Eq.\ \eqref{eq:fisher}, being specifically described at a mesoscopic level by the so-called stochastic FKPP (sFKPP) equation,
\begin{equation}
    \partial_t \rho=D\nabla^2 \rho + \rho - \rho^2 + \sqrt{\rho(1-\rho)/N} \eta(\bm{r},t) ,
    \label{eq:stoch_fisher}
\end{equation}
where $\eta(\bm{r},t)$ is zero-average, uncorrelated Gaussian white noise of unit variance, and $N$ is the number of particles in the system so that Eq.\ \eqref{eq:stoch_fisher} indeed retrieves Eq.\ \eqref{eq:fisher} in the macroscopic $N\to\infty$ limit. Note that alternative microscopic models may yield stochastic generalizations which, while differing from Eq.\ \eqref{eq:stoch_fisher}, still have Eq.\ \eqref{eq:fisher} as a mean-field or macroscopic limit; a celebrated case is directed percolation, see e.g.\ \cite{moro2004,henkel2008}.

The $A+A \leftrightarrow A$ system is one such case in which the hydrodynamic behavior is actually dominated by fluctuations. Indeed, the advancing front displays kinetic roughening  \cite{barabasi1995,krug1997}, namely, scale-invariant fluctuations characterized by critical exponents (see section \ref{observables} below for details) which have been studied systematically
\cite{riordan1995,tripathy2000,moro2001,tripathy2001,moro2004_b}. The main conclusion \cite{moro2001,moro2004_b}, confirmed by recent work on the sFKPP equation \cite{nesic2014}, 
is that the fluctuations of $d_{\parallel}$-dimensional fronts in the ($d_{\parallel}+1$-dimensional) $A+A \leftrightarrow A$ system are in the Kardar-Parisi-Zhang (KPZ) universality class of kinetically rough $d_{\parallel}$-dimensional interfaces. The prime representative of this universality class is the KPZ equation \cite{kardar1986} for the time evolution of the scalar field $h(\bm{r}_{\parallel},t)$ representing the front position, which reads
\begin{equation}
 \partial_t h = \nu \nabla^2 h + \frac{\lambda}{2} (\nabla h)^2 + \eta, \label{eq:kpz}
\end{equation}
where $\bm{r_{\parallel}}\in\mathbb{R}^{d_{\parallel}}$, $\nu>0$ and $\lambda$ are parameters, and the noise term $\eta(\bm{r}_{\parallel},t)$ is zero-average, uncorrelated Gaussian white noise as in Eq.\ \eqref{eq:stoch_fisher}. 

Sparked by exact solutions of this equation and other models in the same universality class for the one-dimensional ($d_{\parallel}=1$) case, the KPZ universality class is quite recently focusing a large attention, see e.g.\ \cite{kriecherbauer2010,halpin-healy2015,Takeuchi2018} for recent reviews. Indeed, these results have shown that KPZ universality goes much beyond the values of the critical exponents. Specifically, the full probability distribution function (PDF) for rescaled field fluctuations is also universal, being (for $d_{\parallel}=1$) a member of the celebrated Tracy-Widom (TW) family of PDF describing the statistics of the largest eigenvalue of random matrices in the Gaussian ensembles \cite{fortin2015}, the precise flavor of the TW distribution depending on global constraints on the system size and/or initial conditions \cite{halpin-healy2015,Takeuchi2018}. Beyond this behavior of the one-point function, analogous strong universality extends even to the two-point function or covariance of front fluctuations, to the extent that currently one-dimensional (1D) fronts in the KPZ universality class are associated with an universal, stationary, stochastic process termed Airy process, which has different variants related with the precise flavor of the TW distribution that occurs in each particular case \cite{prahofer2002,bornemann2008,quastel2011,corwin2013}. These strong universality properties have been fully assessed in experiments \cite{halpin-healy2015,Takeuchi2018} and seem to generalize (albeit in absence of exact results) to higher dimensions ($d_{\parallel}>1$), see e.g.\ \cite{alves2014} and references therein. Overall, KPZ stands out as the prime class for strongly correlated systems displaying universal fluctuations, such as bacterial populations \cite{hallatschek2007}, turbulent liquid crystals \cite{takeuchi2011}, reaction-limited growth \cite{almeida2014}, diffusion-limited growth \cite{nicoli2013}, classical nonlinear oscillators \cite{vanbeijeren2012}, stochastic hydrodynamics \cite{mendl2013}, colloidal aggregation \cite{yunker2013}, random geometry \cite{santalla2015}, superfluidity \cite{altman2015}, active matter \cite{chen2016}, or quantum entanglement \cite{nahum2017}, to cite a few.

While the various traits of 1D KPZ universality have been recently addressed in great detail for interacting particle systems like the totally asymmetric simple exclusion model (TASEP), the polynuclear growth model, etc.\  \cite{kriecherbauer2010,halpin-healy2015,cordoba-torres2018,Takeuchi2018} and for growth models like Eden, ballistic deposition, etc.\ \cite{halpin-healy2015,Takeuchi2018}, reaction-diffusion systems have been comparatively less studied from this point of view. In the case of the $A+A \leftrightarrow A$ system, early numerical work \cite{riordan1995} did obtain kinetic roughening behavior for the dynamics of the front, but characterized by critical exponents with non-KPZ values. Later approaches \cite{tripathy2000} still obtained non-KPZ exponents which were interpreted as evidence for a conjecture \cite{tripathy2001} that $d_{\parallel}$-noisy pulled fronts \cite{saarloos2003} should be in the universality class of the KPZ equation for $(d_{\parallel}+1)$-dimensional interfaces. In turn, more accurate analysis of numerical simulations of the $A+A \leftrightarrow A$ model \cite{moro2001} clarified the situation as being conditioned by the specific method employed for the measurement of the critical exponents, and unambiguously showed that the relevant universality class for the kinetic roughening was indeed that of the KPZ equation for $d_{\parallel}$-dimensional interfaces, in terms of critical exponents. While time-related critical exponent values remained less thoroughly characterized for the discrete model than space-related exponents, more recent work on the sFKPP equation and related continuum systems \cite{moro2004,moro2004_b,nesic2014} has reinforced the consensus on the relevance of KPZ scaling for this class of reaction-diffusion systems. Note that, however, knowledge of critical exponents may not suffice to identify the universality class in kinetic roughening: indeed, examples are known in which, e.g.\ \cite{Saito2012} a linear system (hence, with non-TW one-point statistics) shares the same critical exponent values as the nonlinear 1D KPZ equation, which paradigmatically displays TW one-point fluctuations. 

In this paper we revisit the numerical simulations of the two-dimensional (2D) $A+A \leftrightarrow A$ model in the light of the more recent developments on 1D KPZ universality. Beyond confirmation of space-related \cite{moro2001,nesic2014} and time-related \cite{nesic2014} exponent values (note that results in \cite{nesic2014} are for the sFKPP equation), we address the one and two-point statistics of field fluctuations as a further stringent test of KPZ universality, confirming behavior consistent with the appropriate Airy process. This fully settles the universality class of the 
2D $A+A \leftrightarrow A$ model with respect to the kinetic roughening properties of the front dynamics. Moreover, such a result underscores this class of reaction-diffusion systems as an alternative context for KPZ behavior in terms of e.g.\ potential new experimental realizations, or as a novel point of view on open challenges, such as the properties of this wide universality class in higher dimensions \cite{halpin-healy2015}.

The paper is organized as follows. Section \ref{observables} recalls basic details on the model and simulation procedures and provides the definitions of the quantities that will be measured and contrasted with theoretical expectations. Our numerical results are reported in section  
\ref{sec:results}, which is finally followed by our conclusions and an outlook in section \ref{sec:concl}. Details on the parameter values considered in our simulations and on our statistical data analysis are provided in \ref{details} and \ref{errors}.

\section{Simulation details and observables}
\label{observables}
We study front propagation and fluctuations for $d_{\parallel}=1$, hence we consider a $L_x \times L_y$ lattice, with the front advancing along the $OY$ direction. In the notation introduced below Eq.\ \eqref{eq:fisher}, $\bm{r}=(r_{\parallel},r_{\bot})=(x,y)$. The system sizes $L_x$ and $L_y$ vary within the set of simulations, depending on the particular condition and/or magnitude to be calculated. At $t = 0$, each point of the lattice is occupied by a particle with a probability equal to the equilibrium density $\rho_{eq} = \mu/(1+\mu)$ (see \cite{riordan1995,moro2001} and references therein), so that the initial configuration consists of a number of particles which are uniformly distributed within a region of area $L_x \times L_{y,0}$, with $L_{y,0} < L_y$. Periodic boundary conditions are assumed along the $x$ coordinate.

The time evolution of the particles is ruled by reaction and diffusion. Thus, at a given time, a random particle is chosen. If an adjacent site on the two-dimensional lattice is unoccupied, then either the particle moves to it, with probability $D = 1/4$, or a new particle is created, with probability $\mu D$, at the adjacent site; if the adjacent site is occupied, on the other hand, then the chosen particle is removed, and the particle at the adjacent site remains, with probability $D$. These rules can be schematized as follows,
\medskip

\begin{center}
    $\CIRCLE$ \hspace{0.1cm} $+$ \hspace{0.2cm} $\stackrel{D}{\longleftrightarrow}$ \hspace{0.2cm}  $+$ \hspace{0.1cm} $\CIRCLE$  \hspace{0.4cm}
    \hspace{2cm}
    $\CIRCLE$ \hspace{0.1cm} $+$ \hspace{0.2cm}
    \( \stackrel[\mu D]{D}{\longleftrightarrows}  \)
    \hspace{0.2cm} $\CIRCLE$ \hspace{0.1cm} $\CIRCLE $\\
\end{center}
where a solid circle (plus sign) denotes an occupied (empty) site. According to this, the parameter $\mu$, describing the probability of a creation of a particle, is the birth rate in this process. 

\begin{figure}
    \centering
    \includegraphics[width=1\textwidth]{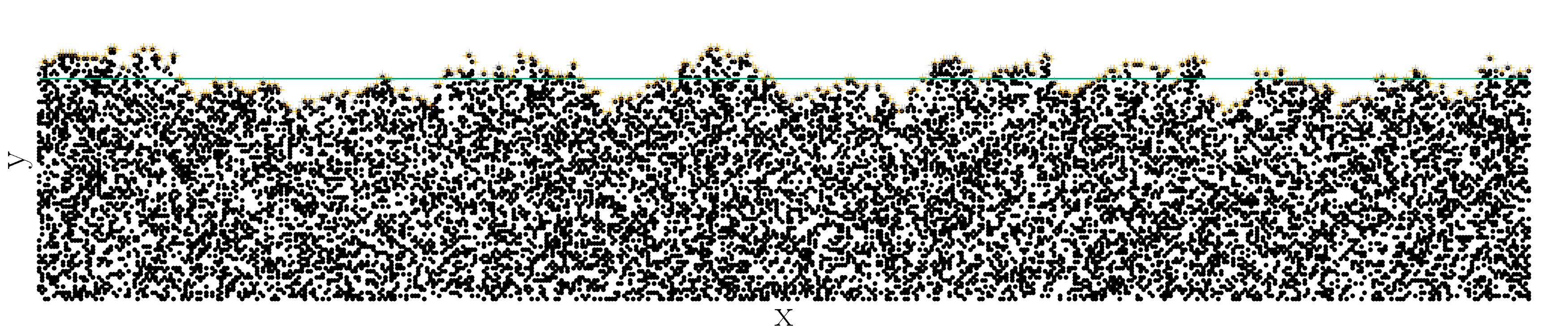}
    \caption{Snapshot of a $L_x=500$ system for $\mu=0.5$ and a fixed value of $t$. The front $h(x,t)$ is plotted as orange crosses and $\bar{h}(t)$ appears as a green line} 
    \label{fig:particles}
\end{figure}

We next describe the different observables which are computed in our numerical simulations. At each time, a local density is defined as
\begin{equation}
    \rho_l(\bm{r},t) = \sum_{\langle \bm{s}, \bm{r} \rangle} n(\bm{s},t),
    \label{eq:local_density}
\end{equation}
where $n(\bm{s},t)$ stands for the occupation of the $\bm{s}$-th site at time $t$, and the sum is over all the nearest neighbors ($\langle \bm{s}, \bm{r}\rangle$) of site $\bm{r}$. Given $\rho_l(\bm{r},t)$, the front position $h(x,t)$ is defined as the maximum $y$-coordinate for sites $\bm{r}=(x,y)$ such that $\rho_l(\bm{r},t) > \rho_{eq}/2$ for the given value of $x$ (and $t$). Alternative related definitions can be employed (see e.g.\ \cite{moro2001}) without relevant changes in the results. For an illustration of our definition, see a sample snapshot of the system from our numerical simulations in Fig.\ \ref{fig:particles}; see \ref{details} for further simulation details. 

The front width (or roughness), $w(L_x,t)$ \cite{barabasi1995,krug1997}, is defined as the standard deviation of the front values,
\begin{equation}
	\label{eq:width}
	w^2(L_x,t)=\left\langle \overline{[h(x,t)-\bar{h}(t)]^2} \right\rangle , 
\end{equation}
where we have used the notation $\overline{O(t)} \equiv (1/L_x)\sum_x O(x,t)$ for a given observable $O(x,t)$ defined on the position of the front. Furthermore, $\langle \cdots \rangle$ denotes average over different realizations of the noise (or initial configurations, or simply runs). Under kinetic roughening conditions, the roughness $w(x,t)$ satisfies the so-called Family-Vicsek scaling law \cite{barabasi1995, krug1997}
\begin{equation}
	\label{eq:w}
	w(L_x,t)=t^{\beta}f\left( t/L_x^z \right),
\end{equation}
in such a way that $w\sim t^{\beta}$ for $t \ll L_x^z$ and $w=w_{\mathrm{sat}}\sim L_x^{\alpha}$ for $t \gg L_x^z$, so that $\alpha=\beta z$. Here, $\alpha$ denotes the so-called roughness exponent, which is related with the fractal dimension of the front \cite{barabasi1995}, while $z$ is the so-called dynamic exponent, which quantifies the power-law increase of the lateral correlation length along the front \cite{barabasi1995,tauber2014},
\begin{equation}
    \label{eq:correlation_length}
    \xi(t) \sim t^{1/z} .
\end{equation}
For later reference, the exact values of these exponents in the one-dimensional KPZ universality class are $\alpha=1/2$, $z=3/2$, and $\beta=1/3$ \cite{barabasi1995,krug1997,kriecherbauer2010,Takeuchi2018}. Moreover, the roughness will be employed to normalize front fluctuations, which will be calculated as
\begin{equation}
	\label{eq:fluctuations}
	\chi(x,t) = \frac {h(x,t) - \bar{h}(t)}{t^{\beta}}\,.
\end{equation}

We define the skewness $S$ and the kurtosis $K$ as functions of the local height fluctuation $\delta h=h(x,t)-\bar{h}(t)$, namely, $S=\langle \delta h^3 \rangle_c / \langle \delta h^2 \rangle_c^{3/2}$ and $K=\langle \delta h^4 \rangle_c / \langle \delta h^2 \rangle_c^{2}$, where $\langle \cdots \rangle_c$ denotes the cumulant average.

Two additional space correlation functions will be considered to describe the spatiotemporal evolution of the front, namely, the height covariance
\begin{equation}
    	C_1(r, t) = \frac 1{L_x} \sum_x \langle h(r+x, t) h(x, t) \rangle-\langle \bar{h}(t)\rangle^2
    \label{eq:correlation_1}
\end{equation}
and the height-difference correlation function
\begin{eqnarray}\nonumber 
	\label{eq:correlation_2}
    	C_2(r, t)&=&\frac 1{L_x} \sum_x \left\langle [h(x+r, t)-h(x, t)]^2 \right\rangle 	  \\
	    &=& 2\langle \overline{h(t)^2}\rangle-\frac 2{L_x}\sum_x \langle h(r+x, t)h(x, t) \rangle,
\end{eqnarray}
where the sum is over all $x$ values. While $C_1(r,t)$ will be used for testing universal properties, $C_2(r,t)$ will allow us to evaluate the correlation length $\xi(t)$. Notice that, again under kinetic roughening conditions \cite{barabasi1995,krug1997}, 
\begin{equation}
    C_2(r,t)=r^{2\alpha} g(r/\xi(t)) ,
    \label{eq:corr_length}
\end{equation}
where $g(u) \sim {\rm const.}$ for $u\ll 1$ and $g(u) \sim u^{-2\alpha}$ for $u\gg 1$. In practice, this allows us to compute the correlation length as e.g.\
\begin{equation}
C_2(\xi_a(t),t)= a \, C_2(L_x/2,t) ,    
\end {equation}
where $a$ is a constant, typically $a=0.8$ or 0.9. The correlation length at a given time $t$ is the distance along the front at which the correlation function $C_2$ takes 80\% or 90\% (respectively) of its plateau-value $C_2(L_x/2,t)$. The precise value of $a$ does not modify the scaling of the the correlation length.

The uncertainties of the fluctuations and the correlation functions have been calculated following the jackknife procedure \cite{peteryoung,Efron1982}; see \ref{errors} for more details.

\section{Results}
\label{sec:results}

In this section we report our results for the front velocity and roughness, the behavior of the correlation length, the universal properties of the front fluctuations, and that of the correlation functions. We refer to the reader to \ref{details} for a complete description of all the runs that we have performed.

\subsection{Front velocity}

\begin{figure}
	\centering
	\includegraphics[width=0.7\textwidth]{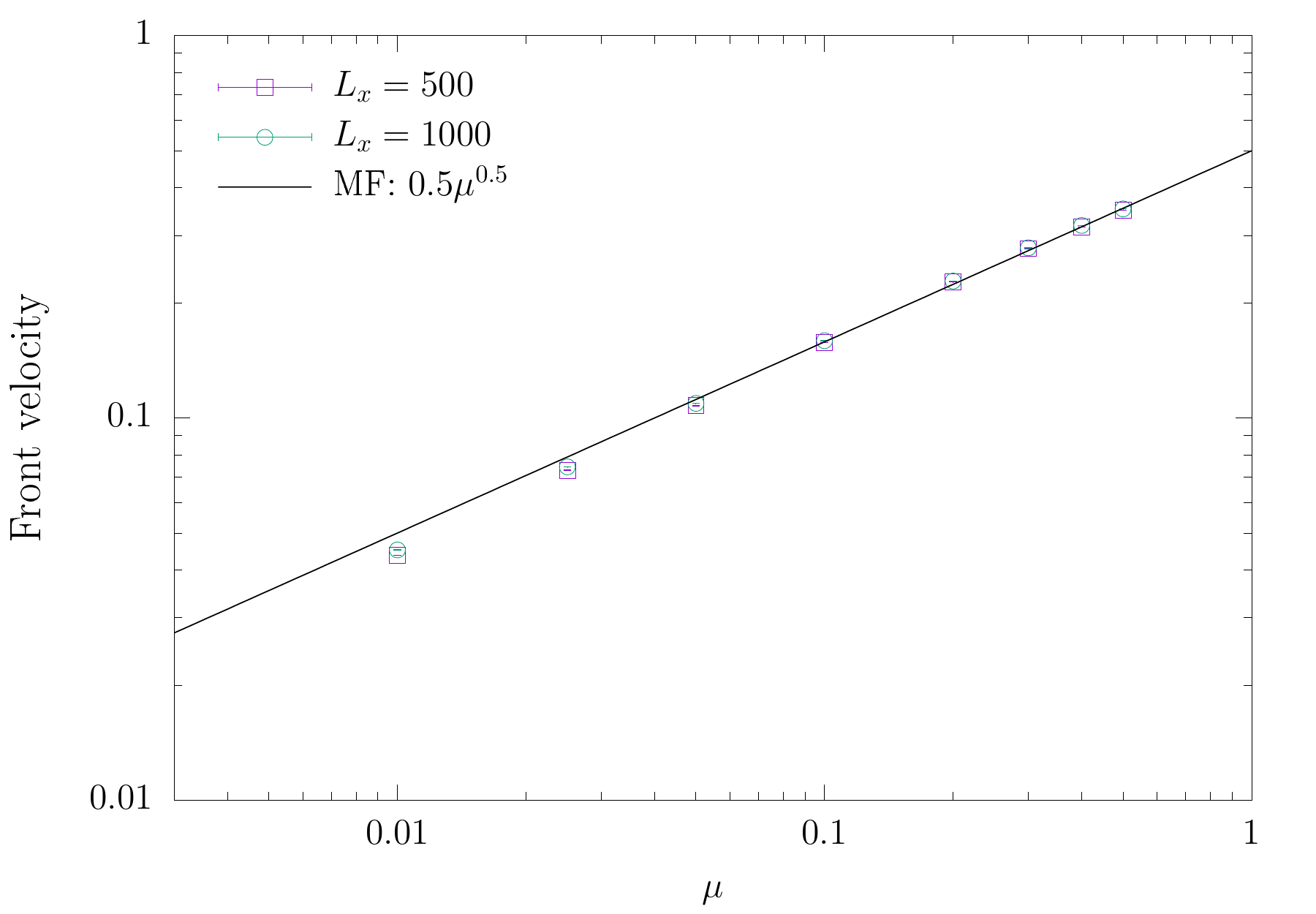}
	\caption{Front velocity versus $\mu$ for two different system sizes (symbols); the error bars are smaller than the symbol sizes. The straight line corresponds to the mean-field prediction. Note that the theoretical prediction has no free parameters.}
	\label{fig:velocity}
\end{figure}

We have computed the mean front position at a given time for two different  system sizes, $L_x = L_y =$ 500 and $L_x=L_y=1000$, using $L_{y,0}=L_y/4$ in both cases. For all simulated values of $\mu$, the mean front position grows linearly with time as $\langle \bar{h}(t) \rangle  = v t+~h_0$. The linear trend of $\langle \bar{h}(t) \rangle $ remains, regardless of the particular $L_x$ and $L_y$ values. Figure \ref{fig:velocity} shows the front velocity $v$ as a function of $\mu$. Within the mean-field (MF) approximation corresponding to Eq.\ \eqref{eq:fisher}, the front velocity $v$ is related to $\mu$ as $v=2D\mu^{1/2}$, where $D$ is the diffusion constant (in our case, $D=1/4$). Simulations show the same behavior, namely, $v~\simeq~0.5\mu^{0.5}$ when $\mu \ge 0.05$, in good agreement with previous results \cite{moro2001}.

\begin{figure}
	\centering
	\includegraphics[width=0.7\textwidth]{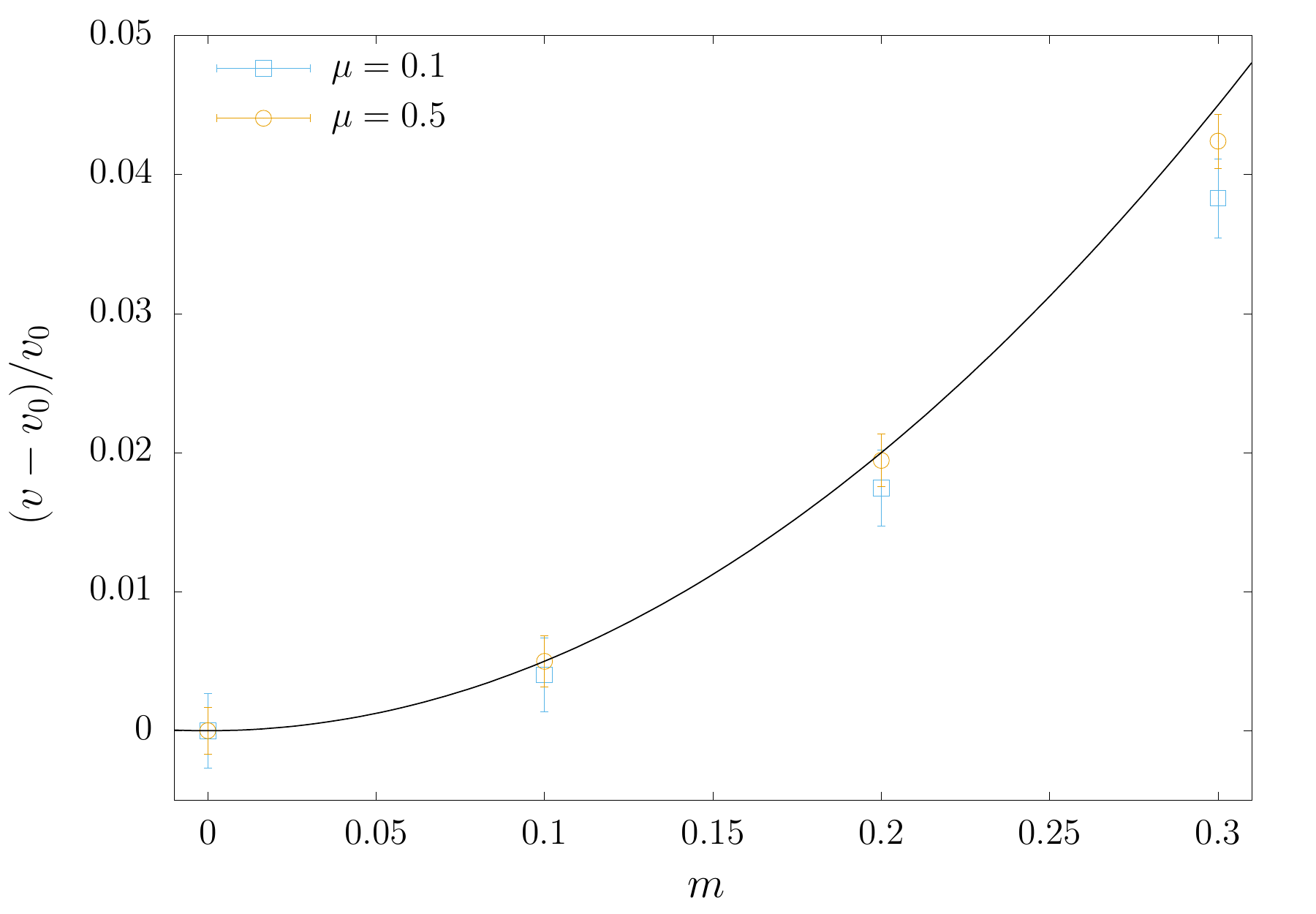}
	\caption{Normalized front velocity, $(v-v_0)/v_0$, versus the average tilt $m$ imposed by helical boundary conditions, where $v_0=v(m=0)$. The system size is $L_x=500$ and two different $\mu$ values are considered (symbols). The solid line shows the $m^2/2$ parabola (no free parameters).}
	\label{fig:tilt}
\end{figure}

To check that the nonlinear term $(\lambda/2)(\nabla h)^2$ of Eq.\ \eqref{eq:kpz} is relevant in the continuum description of our discrete model, we have measured $v$ as a function of the average substrate slope $m$. We have implemented such a slope by introducing helical boundary conditions such that $x(L+1)=x(1)-(L+1)m$. 
The tilt changes the front velocity as $v(m)=v(0)+(\lambda/2)m^2$ \cite{barabasi1995}, with $\lambda=v(0)$ \cite{Pimpinelli1998}.
Our data indeed show the parabolic dependence of the velocity with $m$, see Fig.\ \ref{fig:tilt}. For $\mu=0.1$ and $\mu=0.5$, the values of the coefficient of the nonlinear term are $\lambda=0.1580 \pm 0.0003$ and $\lambda=0.3481 \pm 0.0004$, respectively.

\subsection{Front roughness and growth exponent}

Earlier studies of kinetic roughening in the present $A + A \leftrightarrow A$ model have reported $\beta =0.27(1)$ and $\alpha = 0.41(2)$ \cite{riordan1995,moro2001}, which are close but not equal to the exact values mentioned above for the KPZ universality class of one-dimensional interfaces. As noted in Sec.\ \ref{sec:intro}, this discrepancy between theoretical expectations and computed values has been ascribed  \cite{moro2001} to the imprecise definition of the interface for small $\mu$ values. Indeed, the front position is defined from the local density, see Eq.\ \eqref{eq:local_density}; a small value of $\mu$ generally implies the existence of just a few particles around a given occupied site, which may lead to an underestimate of $\rho_l(\bm{r})$ and, therefore, of the position and width of the interface.

\begin{figure}
	\centering
	\includegraphics[width=0.7\textwidth]{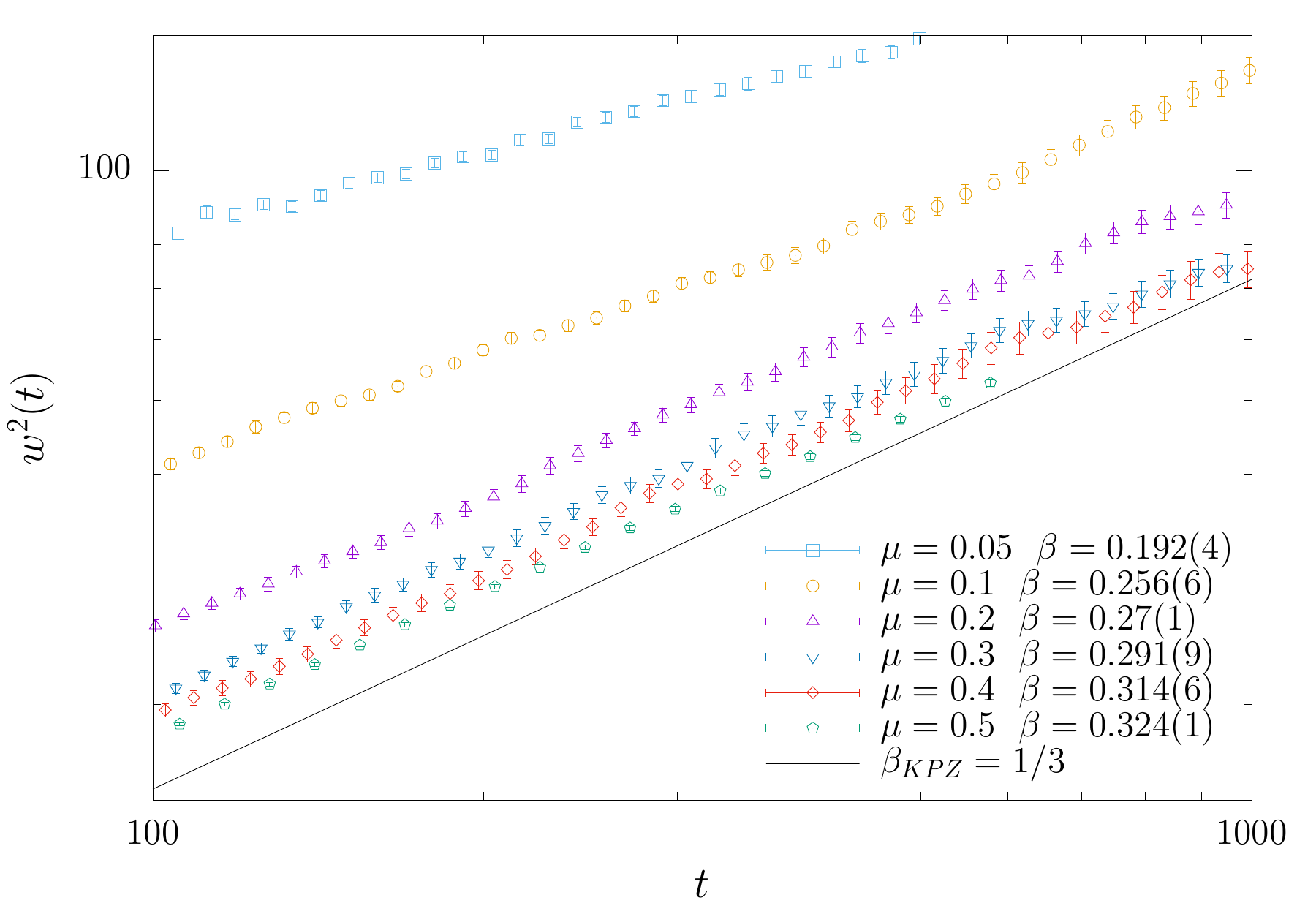} 
	\caption{Squared front roughness, $w^2(t)$, plotted as a function of time for several values of $\mu$ (symbols), as computed for $L_x = L_y=500$. The values of $\beta$ in the legend are obtained as described in the main text. KPZ scaling, $w^2(t) \sim t^{2\beta_{\rm KPZ}} = t^{2/3}$,  corresponds to the solid black line.}
	\label{fig:w_beta}
\end{figure}

This interpretation is confirmed by Fig.\ \ref{fig:w_beta}, which plots the squared front roughness, $w^2(t)$, as a function of time $t$ for $L_x = L_y = 500$ and several values of the birth rate $\mu$. 
Figure~\ref{fig:w_beta} reveals that the roughness exponent $\beta$
depends sensitively on the value of $\mu$, approaching the KPZ limit
for $\mu \simeq 0.5$. As usual, for each data set, the apparent
effective value of $\beta$ characterizing the $w^2(t) \sim t^{2\beta}$
power-law behavior decreases for the longest times, as a symptom of
eventual saturation to the steady-state value of the roughness, due to
the finite size of the simulated systems \cite{barabasi1995,krug1997}.

We have actually computed the front roughness for different system
sizes, see Table \ref{tab:beta}. The statistical errors in the values
of $\beta$ have been calculated by the jackknife method, see
\ref{errors}. We therefore conclude that, as expected, the theoretical
KPZ behavior is systematically approached for increasing $\mu$ and
$L_x$. Exponent values for smaller $\mu$ and $L_x$ are close to
$\beta_{\rm EW}=1/4$, which is the exact value corresponding to the
linearized ($\lambda=0$) KPZ equation, the so-called Edwards-Wilkinson
equation, which frequently provides preasymptotic behavior in the
context of KPZ scaling \cite{barabasi1995,krug1997}. As a consequence
of these numerical results, and unless explicitly indicated, in the
remainder of this paper all our numerical simulations are performed
for $\mu=0.5$.

\begin{table}
    \centering
        \renewcommand{\arraystretch}{1.2}
    \begin{tabular}{m{2cm} m{2cm} m{2cm} m{2cm}}
    \hline \hline
                              & $\mu=0.1$  &  $\mu=0.3$   &        $\mu=0.5$           \\ \cline{1-4}
$L_x=250$   & 0.230(7) & 0.294(6) & 0.314(4) \\ \cline{1-4}
$L_x=500$  & 0.256(6) & 0.291(9) & 0.324(5)   \\ \cline{1-4}
$L_x=1000$ & 0.28(1) & 0.305(4) & 0.320(4) \\ 
\hline \hline
    \end{tabular}
     \caption{Growth exponent $\beta$ for several values of $\mu$ and $L_x$. Note the trend, for higher $L_x$ and $\mu$, towards the KPZ value $\beta_{\mathrm{KPZ}}=1/3$.}
    \label{tab:beta}
\end{table}

\subsection{Height-difference correlation function: dynamic and roughness exponents}

\begin{figure}
\centering
\includegraphics[width=0.7\textwidth]{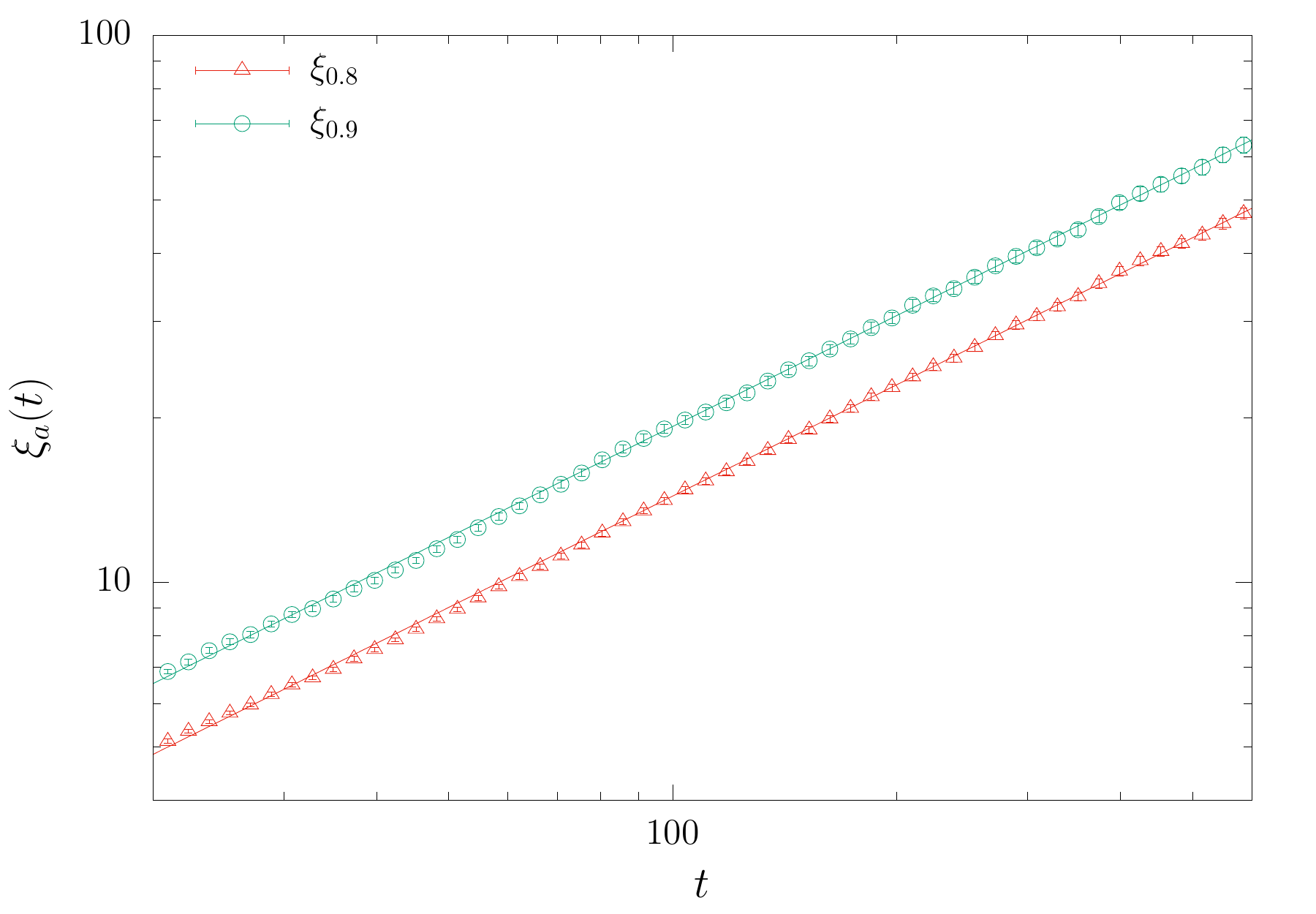}
\caption{Estimates $\xi_{0.8}(t)$ and $\xi_{0.9}(t)$ as functions of time for $\mu=0.5$, and $L_x=500$. The lines correspond to best fits of the numerical values obtained in our simulations (symbols).}
\label{fig:corr_length}
\end{figure}

As described in section \ref{observables}, the correlation length at a given time $t$, $\xi(t)$, can be estimated from the plateau of the $C_2(r,t)$ curves at large enough $r$. We have estimated $\xi_{0.8}$ and $\xi_{0.9}$ as the values of $r$ for which $C_2(r,t)$ equals $0.8P$ and $0.9P$, respectively, where $P=C_2(L_x/2,t)$ is the value of the plateau at time $t$. Figure \ref{fig:corr_length} shows the corresponding estimates of the correlation length as functions of time. 
Fitting these estimates to Eq.\ \eqref{eq:correlation_length} with $t \in [20:600] $, we obtain $1/z=0.673(9)$ from $\xi_{0.9}$ and $1/z=0.676(7)$ from $\xi_{0.8}$, both of them fully compatible (within the error bars) with the exact $1/z_{\rm KPZ}=2/3$ value of the 1D KPZ universality class. Again, the error bars for these values of $1/z$ have been calculated by jackknife, see \ref{errors}. To our knowledge, this is the first direct measurement of the dynamic exponent for the $A+A\leftrightarrow A$ model.

\begin{figure}
\centering
\includegraphics[width=0.7\textwidth]{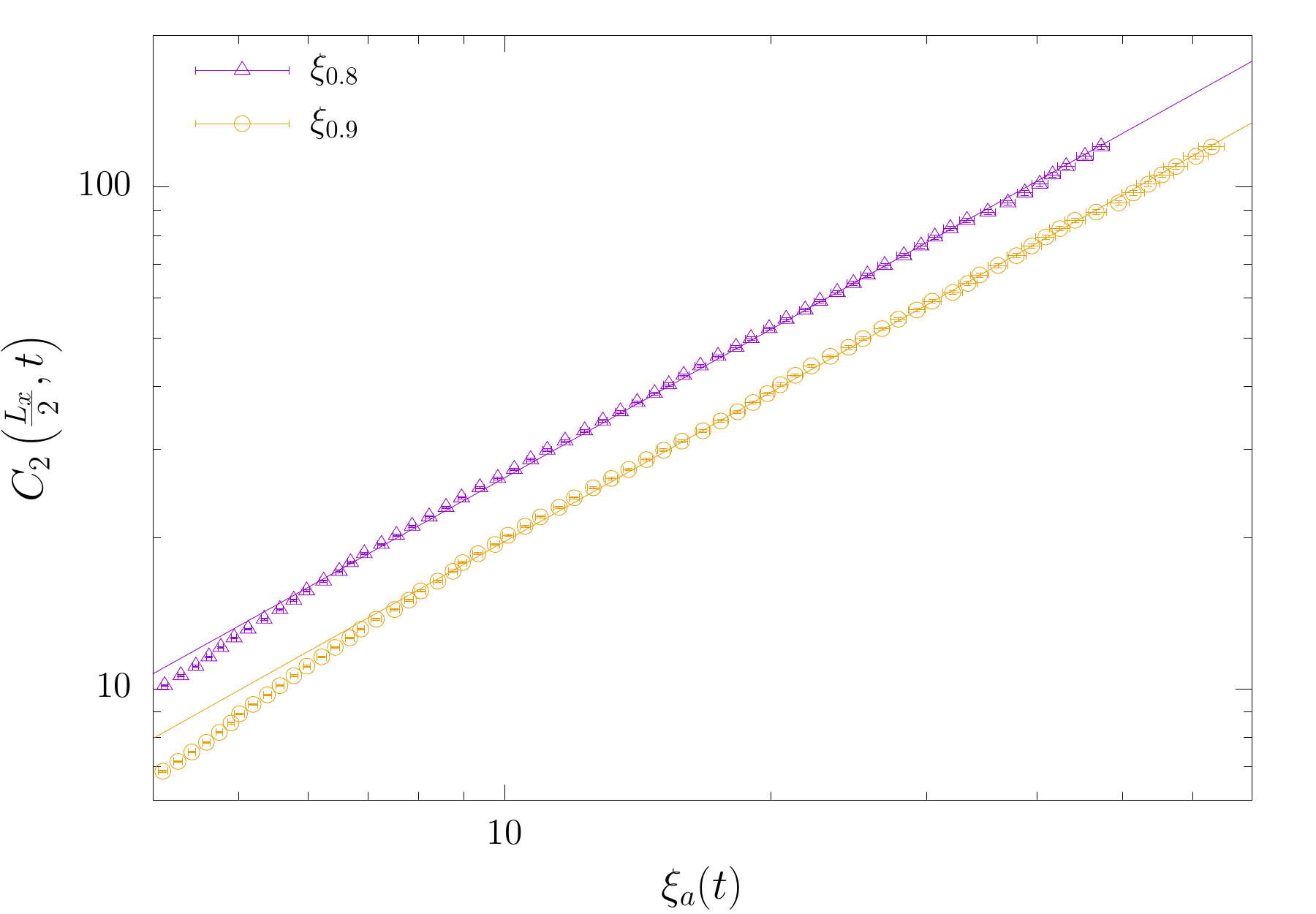}
\caption{Height-difference correlation function $C_2\left(L_x/2,t\right)$ represented against $\xi_{0.8}(t)$ and $\xi_{0.9}(t)$ for different values of time, $\mu=0.5$, and $L_x=500$. The lines are best fits of the numerical values obtained in our simulations (symbols).}
\label{fig:alpha}
 \end{figure}

The height-difference correlation function can also provide the value of the roughness exponent. From Eq.\ \eqref{eq:corr_length}, $C_2(r,t) \sim \xi^{2\alpha}(t)$ for $r \gg \xi(t)$. We have represented $C_2(L_x/2,t)$ against $\xi_{0.8}(t)$ and $\xi_{0.9}(t)$ in order to compute the exponent $\alpha$, see Fig.\ \ref{fig:alpha}, in which best fits have been performed for $t \in [14,600]$ and $t \in [20,600]$, respectively. We thus obtain $2\alpha= 0.994(6)$ for $\xi_{0.8}$ and $2\alpha= 0.99(1)$ for $\xi_{0.9}$, again compatible with the 1D KPZ universality class, namely $2\alpha_{\rm KPZ}=1$.

\begin{figure}
\centering
\includegraphics[width=0.7\textwidth]{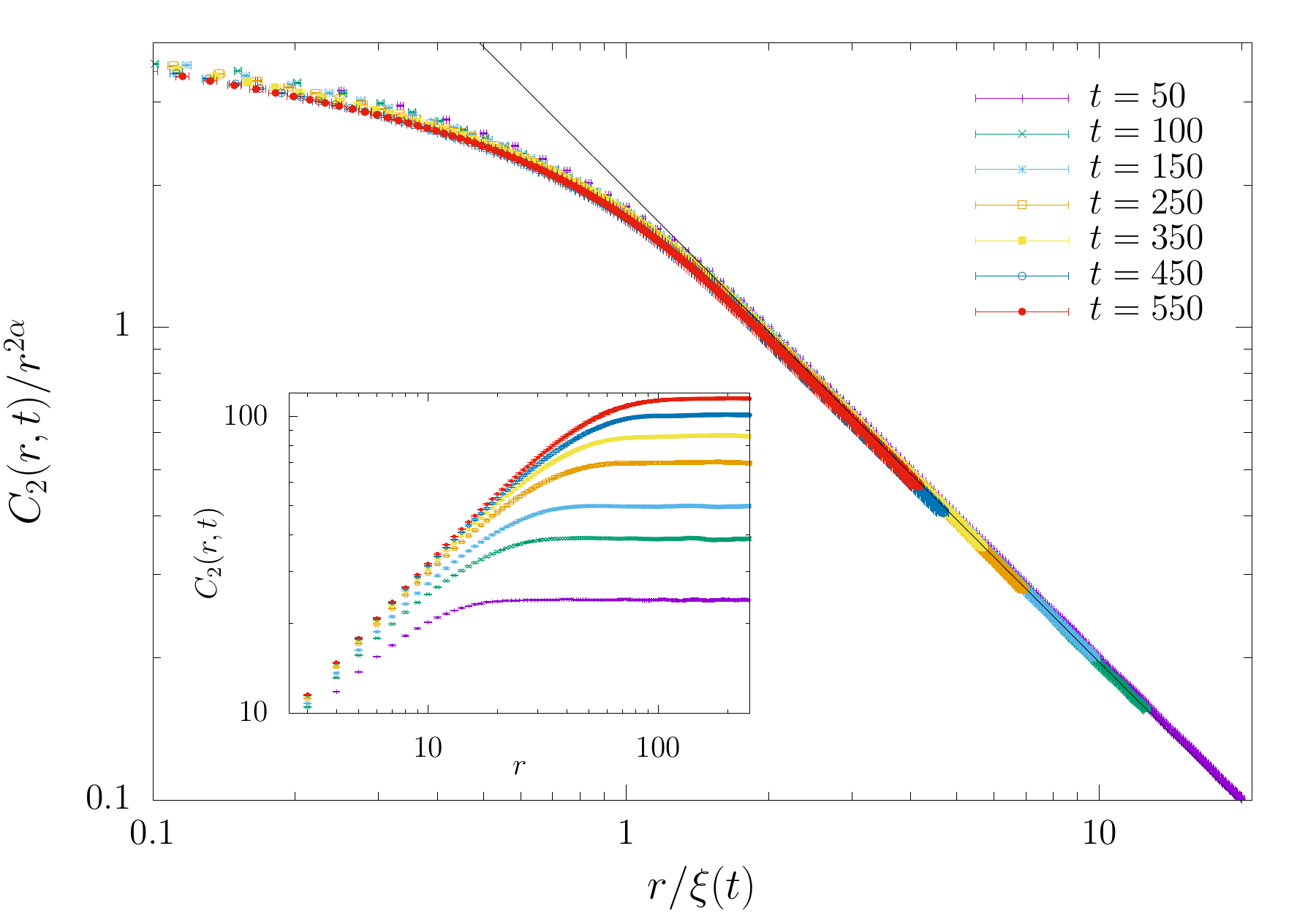}
\caption{Data collapse of the height-difference correlation function obtained for different values of time in numerical simulations for $\mu=0.5$, and $L_x=500$ (symbols), where $\alpha=1/2=\alpha_{\rm KPZ}$ has been assumed. The master curve onto which collapse occurs is the function $g(r/\xi(t))$ of Eq.\ \eqref{eq:corr_length}, the solid black line representing the theoretical behavior for large $u$, $g_{\rm KPZ}(u) \sim u^{-2\alpha_{\rm KPZ}}=u^{-1}$. Inset: The height-difference correlation function $C_2(r,t)$ is shown as a function of $r$ for the same values of time described in the legend of the main panel.}
\label{fig:funcionuniversal}
 \end{figure}

More generally, as a test of Eq.\ \eqref{eq:corr_length} we have represented $C_2(r,t)/r^{2\alpha}$ versus $r/\xi_{0.9}$ assuming $\alpha=0.5$, see Fig.\ \ref{fig:funcionuniversal}. We obtain a time-independent universal function which approximates $g(u)$ in Eq.\ \eqref{eq:corr_length}; both the quality of the collapse and the agreement with the expected universal behavior of the scaling function are better for large values of $r/\xi(t)$.

\subsection{Universality properties of front fluctuations: one-point function}

The one-point statistics of the field fluctuations is known to be another universal trait of the KPZ universality class \cite{kriecherbauer2010,halpin-healy2015}. Since we are employing periodic boundary conditions for 1D interfaces, the PDF of rescaled front fluctuations, Eq.\ \eqref{eq:fluctuations}, should be provided by the Tracy-Widom distribution for the largest eigenvalue of a random-matrix in the Gaussian orthogonal ensemble (TW-GOE) \cite{kriecherbauer2010,halpin-healy2015}.

\begin{figure}
    \centering
    \includegraphics[width=0.7\textwidth]{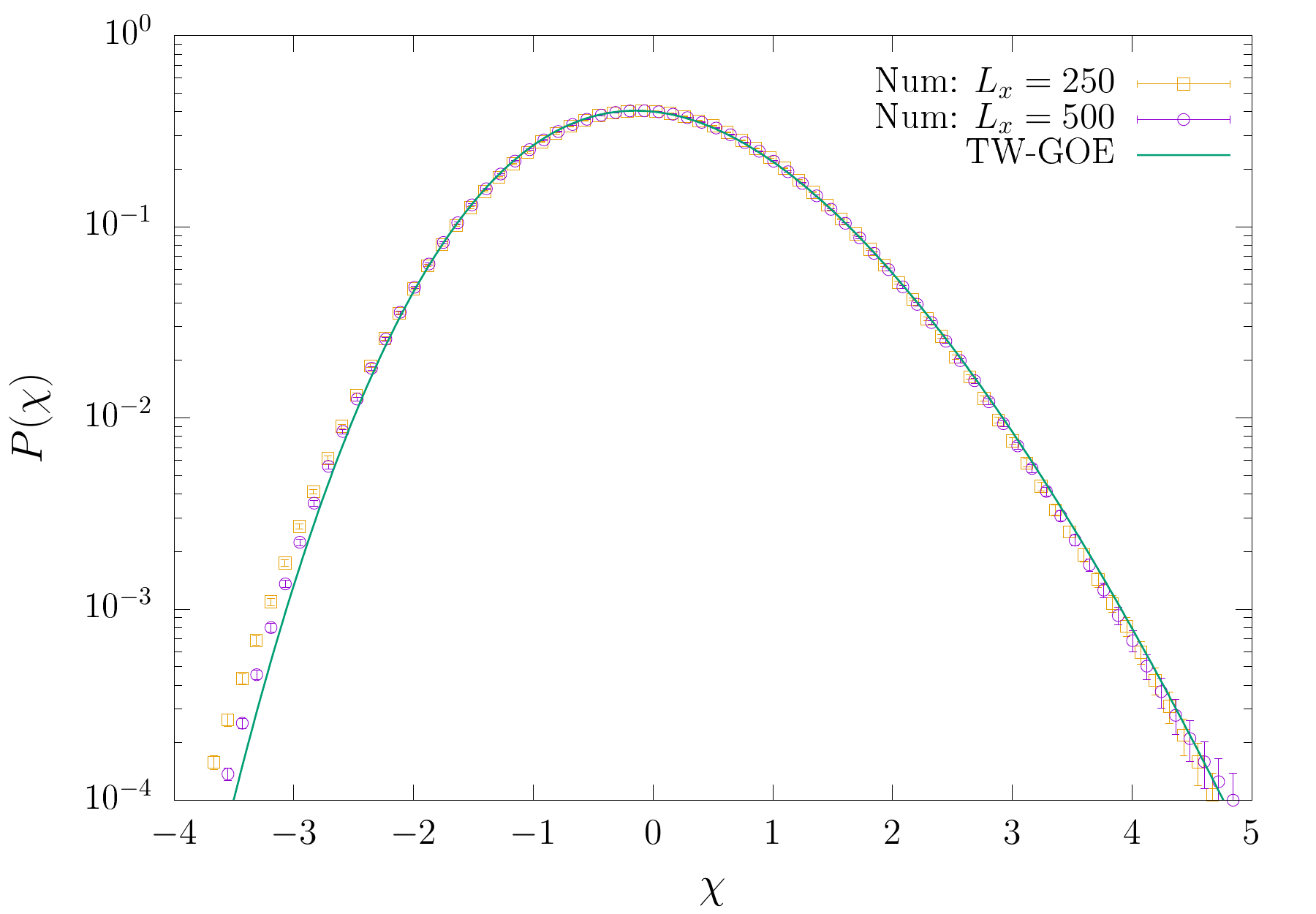} 
    \caption{Histograms of front fluctuations for different $L_x$ and $\mu = 0.5$. The TW-GOE theoretical prediction is shown as a solid line. For  $L_x=250$ ($L_x=500$) we have used times within the interval $t\in [10,300]$ ($t\in [120,600]$).
    } 
    \label{fig:tw_Lx}
\end{figure}

Figure \ref{fig:tw_Lx} plots the histogram of front fluctuations computed for $L_x =250$ and $L_x=500$, $\mu = 0.5$, and $t \ll L_x^z$ (that is, before  front saturation to steady state), together with the exact TW-GOE distribution (solid line). We have also computed the skewness $S$ and the kurtosis $K$ for the numerical distributions. We have obtained $S=0.22 \pm 0.01$, $K=0.16 \pm 0.02$ for $L_x=250$ and $S=0.25 \pm 0.01$, $K=0.17\pm 0.03$ for $L_x=500$. The numerical values for the TW-GOE PDF are $S=0.29346452408$ and $K=0.1652429384$ \cite{Bornemann2010_2}.
Clearly our data compare better with the theoretical expectations for larger $L_x$, hence we interpret the small differences between the exact TW-GOE values and those characterizing our data as due to the finite size of our simulated systems.

\subsection{Universality properties of front fluctuations: height covariance}

\begin{figure}
\centering
\includegraphics[width=0.7\textwidth]{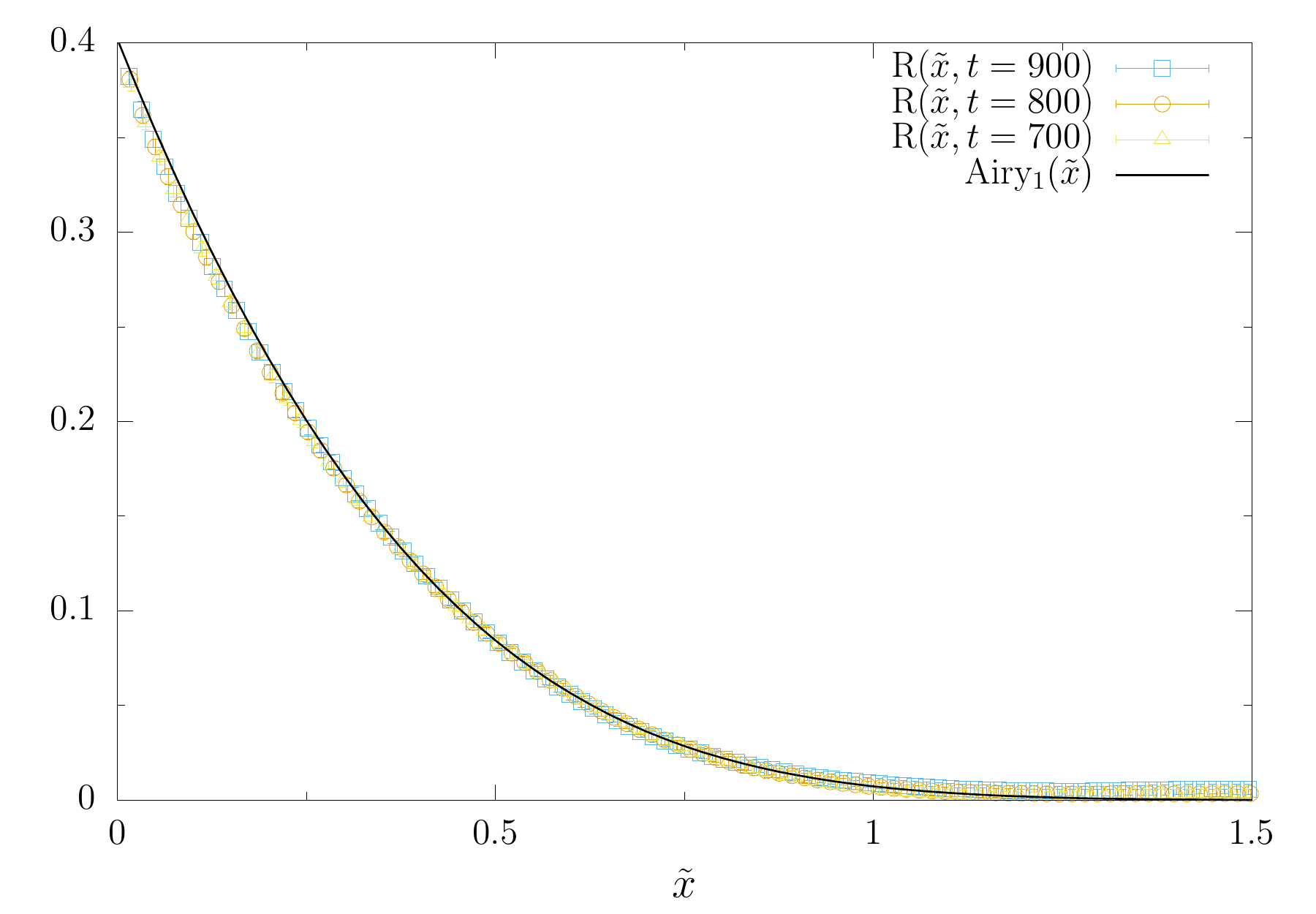}
\caption{The solid line shows the $\mathrm{Airy}_1(\tilde{x})$ function. The values of the numerical data from our simulations (symbols) are proportional to $C_1(r,t)$ for $t=700$, 800, and 900, as obtained for a system with $L_x=L_y=512$ and $\mu=0.5$. In particular, we represent $R(\tilde{x},t) \equiv C_1 \left(\tilde{x} t^{2/3}/a_2\right)/(a_1 t^{2/3})$ 
with $a_1=2.15$ and $a_2=0.74$, see Eqs.\ \eqref{eq:airy1} and \eqref{eq:airy1b}. To improve visibility, the figure only represents every second point.}
\label{fig:airy1}
 \end{figure}

Universal behavior in the 1D KPZ equation occurs for additional magnitudes \cite{halpin-healy2015,Takeuchi2018}, like the full space-time behavior of the height covariance $C_1(r,t)$, Eq.\ \eqref{eq:correlation_1}. Indeed, under periodic boundary conditions, this is expected to behave as
\begin{equation}
    C_1(r,t)=a_1 \, t^{2\beta} \mathrm{Airy}_1\left(a_2 r/t^{1/z} \right) ,
   \label{eq:airy1}
\end{equation}
where $\mathrm{Airy}_1(u)$ denotes the covariance of the Airy$_1$ process
\cite{bornemann2010,halpin-healy2015,Takeuchi2018}, and $a_1$ and $a_2$ are suitable numerical constants \cite{alves2011,Oliveira2012,nicoli2013}
which need to be estimated in order to test Eq.\ \eqref{eq:airy1} for our simulations. The value of $a_1$ is given by
\begin{equation}
a_1=\frac{C_1(0,t)}{t^{2/3}\mathrm{Airy}_1(0)} .
\label{eq:a1}
\end{equation}
We can then estimate the value of $a_2$ by choosing a point of the graph of the Airy$_1$ function, ($\tilde{x}$, Airy$_1(\tilde{x})$). Specifically, in our analysis we have selected $\tilde{x}=\tilde{x}_0=0.5$. The relation between $\tilde{x}$ and $a_2$ is $\tilde{x}\equiv a_2 r/t^{2/3}$. Then, from Eq.\ \eqref{eq:airy1},
\begin{equation}
\frac{C_1\left(\tilde{x}_0 t^{2/3}/a_2\right)
}{t^{2/3}}=a_1\mathrm{Airy}_1\left( \tilde{x}_0 \right) .
\label{eq:airy1b}
\end{equation}
We know the value of $C_1\left( \tilde{x}_0 t^{2/3}/a_2 \right)$ and, by linear polynomial interpolation, with our data we can calculate the value of its argument and solve for $a_2$. We have tested this scaling form and obtained a very good scaling plot, see Fig.\ \ref{fig:airy1}.

\section{Conclusions and outlook}
\label{sec:concl}

In this work we have revisited the 2D $A+A \leftrightarrow A$ reaction-diffusion model via numerical simulations, in order to fully assess the kinetic roughening behavior of the evolving front which ensues, from the point of view of the 1D KPZ universality class. Beyond confirming the numerical values of the critical exponents, we have elucidated explicitly the one- and two-point statistics as corresponding to the Airy$_1$ process, as expected for our present choice of periodic boundary conditions. 
Actually, for systems in the 1D KPZ class, the dynamical behavior is known to be particularly rich and complex, including a number of additional,  interesting properties, such as ergodicity loss and aging, non-trivial persistence, peculiar fluctuation properties around steady steady state, etc.\ \cite{halpin-healy2015,Takeuchi2018}. Such properties might warrant further detailed study in future for reaction-diffusion systems of the type that we have addressed here. 

In our present work, we have considered exponents and one- and two-point statistics as the main traits characterizing the universality class, as is currently being done in the context of kinetic roughening \cite{halpin-healy2015,Takeuchi2018}. Indeed, identification of surface kinetic roughening universality classes, taking into account additional properties beyond exponent values, is becoming increasingly pertinent in view of potential ambiguities \cite{Saito2012} and, more generally, because it provides an improved understanding of scale invariance far-from-equilibrium, not only in the KPZ case, but in other universality classes as well \cite{nicoli2013,Carrasco2016,Rodriguez-Fernandez2019}. Specifically, in our system we have confirmed that the 2D $A+A \leftrightarrow A$ model coincides with the stochastic FKPP equation, Eq.\ \eqref{eq:stoch_fisher}, for $\mathbf{r}\in\mathbb{R}^2$, with respect to the the full set of scaling exponents as determined in \cite{moro2004,moro2004_b,nesic2014}, and with respect to the one-point statistics as preliminarily obtained in \cite{nesic2015}. Beyond this, our present study confirms explicitly the expected behavior of the field two-points statistics in terms of the Airy$_1$ covariance.

From a more general point of view, our study characterizes a peculiar type of fluctuations which may likely be found within the wide class of physical and biological systems; these are described at a mean-field level by the Fisher (or, more properly, the FKPP) equation, Eq.\ \eqref{eq:fisher}. For instance, recent advances in (bio)materials science are enabling material design and control at submicron and nano scales \cite{Scalise2019} through DNA circuits based on reaction-diffusion systems \cite{Zenk2017,Lovrak2019}. Working at such small scales, it is important to assess the potential quantitative and qualitative influence of external and internal noise in the relevant reaction-diffusion processes and systems. At this, the $A+A \leftrightarrow A$ model is a prime example of cases in which microscopic fluctuations can have macroscopic implications \cite{nesic2014}. Interestingly, and keeping within KPZ-related interacting particle systems, we recall that the paradigmatic asymmetric simple exclusion process (ASEP) model was historically put forward in the 1960's as a simplified description of the dynamics of ribosomes translating along a messenger RNA molecule \cite{Chou2011}. Recent results on 1D KPZ statistics \cite{halpin-healy2015,Takeuchi2018} are elucidating fluctuation properties which perhaps await to be found in biophysics and cellular biology at the few-molecules level.

\section{Acknowledgments}
We thank E.\ Moro and S.\ Nesic for interaction and exchange. This work was partially supported by Ministerio de Econom\'{\i}a y Competitividad, Agencia  Estatal  de  Investigaci\'on, and Fondo Europeo  de  Desarrollo  Regional  (FEDER) (Spain and  European Union) through grants No.\ FIS2016-76359-P, No.\ FIS2015-66020-C2-1-P, and No.\ PGC2018-094763-B-I00, and by Junta de Extremadura (Spain) through grants No.\ GRU10158 and No.\ IB16013 (partially funded by FEDER). We have run our simulations in the computing facilities of the Instituto de Computaci\'{o}n Cient\'{\i}fica Avanzada de Extremadura (ICCAEx).

\section*{References}

\bibliographystyle{iopart-num}

\appendix
\section{Simulation parameters} 
\label{details}
In this Appendix we collect all parameter details for the numerical simulations reported in the paper. Specifically, Tables \ref{tab:det} and \ref{tab:det2} record all the simulation conditions that we have considered. Note that the time step in the simulation is taken as the value required for all the particles to have the chance to diffuse, hence $\delta t=1/N(t)$, where $N(t)$ is the total number of particles at time $t$. 

\begin{table} [h]
\begin{center}
\begin{tabular}{|c|c|c|P{1.5cm}|P{2cm}|P{1.5cm}|}
\hline
$L_x$ & $L_y$ & $L_{y,0}$ & $\mu$ & $t_\mathrm{max}$ & runs \\ \hline
 &  & & 0.1 & $7.5 \times 10^6$ & 104 \\ \cline{4 -6}
250 & 1000 & 250 & 0.3 & $7.5\times 10^6$ & 100 \\ \cline{ 4- 6}
 &  &  & 0.5 & $7.5\times 10^6$ & 1600 \\ \hline

 & & & 0.01 & $1.1\times 10^4$ & 100 \\ \cline{ 4- 6}
 & &  & 0.025 & $5.8\times 10^3$ & 100 \\ \cline{ 4 - 6}
&  &  & 0.05 & $3.8\times 10^3$ & 100 \\ \cline{ 4- 6}
500 & 500 & 125  & 0.1 & $2.6\times 10^3$ & 100 \\ \cline{ 4- 6}
 & & & 0.2 & $1.9\times 10^3$ & 100 \\ \cline{ 4- 6}
 &  &  & 0.3 & $1.5\times 10^3$ & 100 \\ \cline{ 4- 6}
 &  &  & 0.4 & $1.2\times 10^3$ & 100 \\ \cline{ 4- 6}
 &  &  & 0.5 & $6.0\times 10^2$ & 1600 \\ \hline

512 & 512 & 128 & 0.5 & $9.3 \times 10^2$ & 2002 \\ \hline

 & & & 0.01 & $2.0\times 10^4$ & 100 \\ \cline{ 4- 6}
& & & 0.025 & $1.0\times 10^4$ & 100 \\ \cline{ 4- 6}
 & & & 0.05 & $7.8\times 10^3$ & 100 \\ \cline{ 4- 6}
1000 & 1000 & 250 & 0.1 & $5.4\times 10^3$ & 100 \\ \cline{ 4- 6}
 & & & 0.2 & $3.7\times 10^3$ & 100 \\ \cline{ 4- 6}
 & & & 0.3 & $2.8\times 10^3$ & 100 \\ \cline{ 4- 6}
 & & & 0.4 & $2.5\times 10^3$ & 100 \\ \cline{ 4- 6} 
 & & & 0.5 & $2.0\times 10^3$ & 100 \\ \hline
\end{tabular}
\end{center}
    \caption{Parameter values for simulations employing periodic boundary conditions. Here, $L_x$ and $L_y$ are the dimensions of the simulation lattice, while $L_{y,0}$ indicates the size along the $y$-direction of the region in which particles are homogeneously distributed at $t=0$, and $t_\mathrm{max}$ is the maximum time that is reached in the simulations. The last column shows the number of runs performed in each case.}
    \label{tab:det}
\end{table}

\begin{table} [h]
\begin{center}
\begin{tabular}{|c|c|c|c|P{1.5cm}|P{2cm}|P{1.5cm}|}
\hline
$L_x$ & $L_y$ & $L_{y,0}$ & $\mu$ & $m$ & $t_\mathrm{max}$ & runs \\ \hline
 & & & & 0 & $1.8\times 10^3$ & 200 \\ \cline{5 -7}
 & & & 0.1 & 0.1 & $1.8\times 10^3$ & 200 \\ \cline{ 5- 7}
 & & & & 0.2 & $1.7\times 10^3$ & 200 \\ \cline{ 5- 7}
 500 & 2000 & 125 & & 0.3 & $1.6\times 10^3$ & 200 \\ \cline {4-7}
 & & & & 0 & $6.0 \times 10^2$ & 200 \\ \cline{5 -7}
 & & & 0.5 & 0.1 & $5.6 \times 10^2$ & 200 \\ \cline{ 5- 7}
 & & & & 0.2 & $5.2 \times 10^2$ & 200 \\ \cline{ 5- 7}
 & & & & 0.3 & $4.8 \times 10^2$ & 200 \\ \hline

\end{tabular}
\end{center}
    \caption{Parameter values for simulations employing helical boundary conditions with an overall slope $m$. Parameters $L_x$ and $L_y$ are as in Table \ref{tab:det}. The initial configurations occupy a trapezoid area with $L_{y,0}$ and $L_{y,0}+mL_x$ heights and $t_{\mathrm{max}}$ is the maximum time that is reached in the simulations. The last column shows the number of runs performed.}
    \label{tab:det2}
\end{table}

\section{Details of the analysis of the data} \label{errors}

In this Appendix we describe the methodology employed for our statistical data analysis. We have computed the statistical error on different observables for highly-correlated rough numerical data. In order to do that, we have followed the method described in Ref.\ \cite{Yllanes2011}, see also \cite{Michael1994,Lulli2016,Seibert1994}.

To study a certain system we perform several trials. As particles are created randomly, times between runs are not the same. To compare a magnitude we define temporal boxes (with a width $\Delta t$) in which we include all the points of the different simulations corresponding to time $t \in (t, t+\Delta t)$. We consider the best estimate of a magnitude $x$ in the temporal box $(t, t+\Delta t)$ of the $i$-th run as the sample mean of the all the points, namely,
\begin{equation}
\hat{x}_i=\frac{1}{n} \sum_{j=1}^n x_j \,,
\end{equation}
where $n$ is the number of points included in that time interval. The mean, $\bar{x}$, is in turn given by
\begin{equation}
  \bar{x}_i=\frac{1}{N} \sum_{i=1}^N \hat{x}_i \,,
\end{equation}
where $N$ is the number of runs (initial conditions).

As a general rule, the errors for all the results reported in the text have been calculated with the jackknife procedure \cite{peteryoung, Efron1982}. The $i$-th jackknife estimate of a magnitude $x$ is the average over all the runs, but omitting the data for the $i$-th run:
\begin{equation}
x_i^{\mathrm{JK}}=\frac{1}{N-1}\sum_{k=1, k \neq i}^N \hat{x}_k\,.
\end{equation}
The variance of $\bar{x}$ is then:
\begin{equation}
    \sigma_\mathrm{JK}(\bar{x})=\frac{N-1}{N} \sum_{k=1}^N(\bar{x}-x_i^\mathrm{JK})^2\,. 
\end{equation}
Hence, for each temporal box we have the estimate $\bar{x} \pm \sqrt{\sigma_\mathrm{JK}}$ (one standard deviation).

To determine a given critical exponent we need to do a fitting in time. It is very important to realize that the data used in a typical fit show a huge correlation (e.g., the $\xi(t) \propto t^{1/z}$ data are highly correlated among them). Hence, one should use the full covariance matrix to perform the global fit in order to compute the given exponent. In general, the full covariance matrix is singular or almost singular (i.e., its determinant is close to zero) \cite{Yllanes2011, Michael1994,Lulli2016,Seibert1994}, which prevents the computation of the matrix inverse required for the fitting procedure. In order to circumvent this problem, we have used the following procedure, which takes into account the statistical correlation of the data and has demonstrated excellent performance e.g.\ in the study of spin glasses \cite{Yllanes2011,Lulli2016}:
\begin{enumerate}
    \item  The mean value, $\bar{z}$, of a given exponent (e.g., $z$) is computed using the data from all the runs using the diagonal covariance matrix of the data.
    \item The statistical error for this exponent is computed using the jackknife procedure. We remove the $i$-th run from the data and compute the $i$-th value, $z_i^\mathrm{JK}$, associated to this jackknife block, using again the its diagonal covariance matrix.
    The error is computed using the standard equation in the jackknife procedure, namely,
   \begin{equation}
    \sigma_\mathrm{JK}(\bar{z})=\frac{N-1}{N} \sum_{k=1}^N(\bar{z}-z_i^\mathrm{JK})^2\,. 
\end{equation} 
\end{enumerate}

Notice that if we use only the diagonal covariance matrix, the statistical error of the exponent will be strongly underestimated (for instance, using the fitting procedure of Gnuplot). With the procedure presented here, we take into account the strong correlation among the data, and provide the exponent with the right statistical error.

Finally, note that we have chosen the time intervals of all the fits in order to obtain $\chi^2/\mathrm{d.o.f.} \approx 1$ \cite{peteryoung}, where $\chi^2$ has been  computed assuming a diagonal covariance matrix and d.o.f. is the number of deegres of freedom of the fit. 

\end{document}